\documentclass[10pt,conference]{IEEEtran}
\AtBeginDocument{%
  \providecommand\BibTeX{{%
    \normalfont B\kern-0.5em{\scshape i\kern-0.25em b}\kern-0.8em\TeX}}}
    
\IEEEoverridecommandlockouts
\usepackage{cite}
\usepackage{amsmath,amssymb,amsfonts}
\def\BibTeX{{\rm B\kern-.05em{\sc i\kern-.025em b}\kern-.08em
    T\kern-.1667em\lower.7ex\hbox{E}\kern-.125emX}}
\usepackage{url}
\usepackage{balance,booktabs,stmaryrd,listings,multirow,wrapfig,graphicx,caption,dashbox,pifont,xcolor,colortbl,color,soul,enumitem}
\definecolor[named]{ACMPurple}{cmyk}{0.55,1,0,0.15}
\definecolor[named]{ACMDarkBlue}{cmyk}{1,0.58,0,0.21}
\usepackage[hidelinks,colorlinks=true,linkcolor=ACMPurple,citecolor=ACMPurple,urlcolor=ACMDarkBlue]{hyperref}
\usepackage{semantic}
\usepackage{lscape}
\usepackage{wasysym}
\usepackage{tikz}
\usepackage{setspace}
\usepackage{booktabs}
\usepackage{bm}
\usepackage{subcaption}
\usepackage[ruled,linesnumbered,vlined]{algorithm2e}
\usetikzlibrary{fit,calc}
\usepackage{tcolorbox}

\usepackage{amsthm}
\theoremstyle{definition}
\newtheorem{definition}{Definition}
\newtheorem{theorem}{Theorem}

\newtheorem{example}{Example}

\usepackage[utf8]{inputenc}
\usepackage{cleveref}
\crefname{section}{§}{§§}
\Crefname{section}{§}{§§}

\usepackage{xcolor}

\usepackage{tcolorbox}

\usepackage{longtable}
\usepackage{lipsum} 

\usepackage{color}

\begin{document}

\title{Boosting Path-Sensitive Value Flow Analysis via \\ Removal of Redundant Summaries}

\author{\IEEEauthorblockN{Yongchao Wang, Yuandao Cai,
and Charles Zhang}
\IEEEauthorblockA{
Department of Computer Science and Engineering \\
The Hong Kong University of Science and Technology, Hong Kong, China}
Email: \{ywanghz, ycaibb, charlesz\}@cse.ust.hk 
}










\maketitle

\begin{abstract}
  Value flow analysis that tracks the flow of values via data dependence is a widely used technique for detecting a broad spectrum of software bugs.
However, the scalability issue often deteriorates when high precision (i.e., path-sensitivity) is required, as the instantiation of function summaries becomes excessively time- and memory-intensive.
The primary culprit, as we observe, is the existence of redundant computations resulting from blindly computing summaries for a function, irrespective of whether they are related to bugs being checked.
To address this problem, we present the first approach that can effectively identify and eliminate redundant summaries, thereby reducing the size of collected summaries from callee functions without compromising soundness or efficiency.
Our evaluation on large programs demonstrates that our identification algorithm can significantly reduce the time and memory overhead of the state-of-the-art value flow analysis by 45\% and 27\%, respectively. 
Furthermore, the identification algorithm demonstrates remarkable efficiency by identifying nearly 80\% of redundant summaries while incurring a minimal additional overhead.
In the largest \textit{mysqld} project, the identification algorithm reduces the time by 8107 seconds (2.25 hours) with a mere 17.31 seconds of additional overhead, leading to a ratio of time savings to paid overhead (i.e., performance gain) of 468.48 $\times$. 
In total, our method attains an average performance gain of 632.1 $\times$.
\end{abstract}


\section{Introduction}
\label{sec:intro}

Path-sensitive value-flow analysis~\cite{cherem2007practical,livshits2003tracking,shi2018pinpoint,fan2019smoke, shi2020conquering,shi2021path, sui2014detecting, sui2016sparse, sui2016svf, wang2023anchor, wang2022complexity} is highly effective in detecting a broad spectrum of software bugs,
such as memory leaks in resource usage, null pointer dereference in memory safety, and the propagation of tainted data in security properties,
by tracking the flow of values along data dependence relations.
Essentially, detecting these bugs boils down to collecting feasible source-sink paths over a program dependence graph~\cite{cherem2007practical, shi2018pinpoint}.
For instance, detecting the null pointer dereference (NPD), considering the null value as the source and the pointer dereference statement as the sink. 
The process involves a two-step process: collecting paths that link a null value and a pointer dereference statement, and then verifying the satisfiability of the path conditions for those paths.

To scale the analysis to large-scale software systems with millions of lines of code, existing approaches~\cite{shi2018pinpoint, shi2020conquering, shi2021path, fan2019smoke, babic2008calysto, xie2005scalable,wu2024libalchemy} employ a bottom-up strategy to gather feasible source-sink paths. 
Specifically, when analyzing a function, these approaches compute the intra-procedural value-flow paths and the corresponding path conditions as function summaries.
The value-flow paths and conditions are referred to as summary paths and summary conditions, respectively.
To ensure that only feasible summaries are collected, a constraint solver is invoked to verify the summary condition once the summary path is collected, despite being a computationally costly process.
To avoid redundant path searching and analysis of callee functions, 
existing approaches clone the summaries of callees and reuse them continuously to supplement the more extended summaries collected within caller functions. 
The summary cloning and summary condition verification process continues until the highest function in the call graph (known as the root) is reached. 
At this point, the algorithm can directly identify source-sink paths by examining summary paths originating from sources and terminating at sinks. 
Since each summary path carries its corresponding path conditions, we can use the terms ``summary'' and ``summary path'' interchangeably without losing generality.

\begin{figure*}[t]
    \centering
    \includegraphics[width=\textwidth]{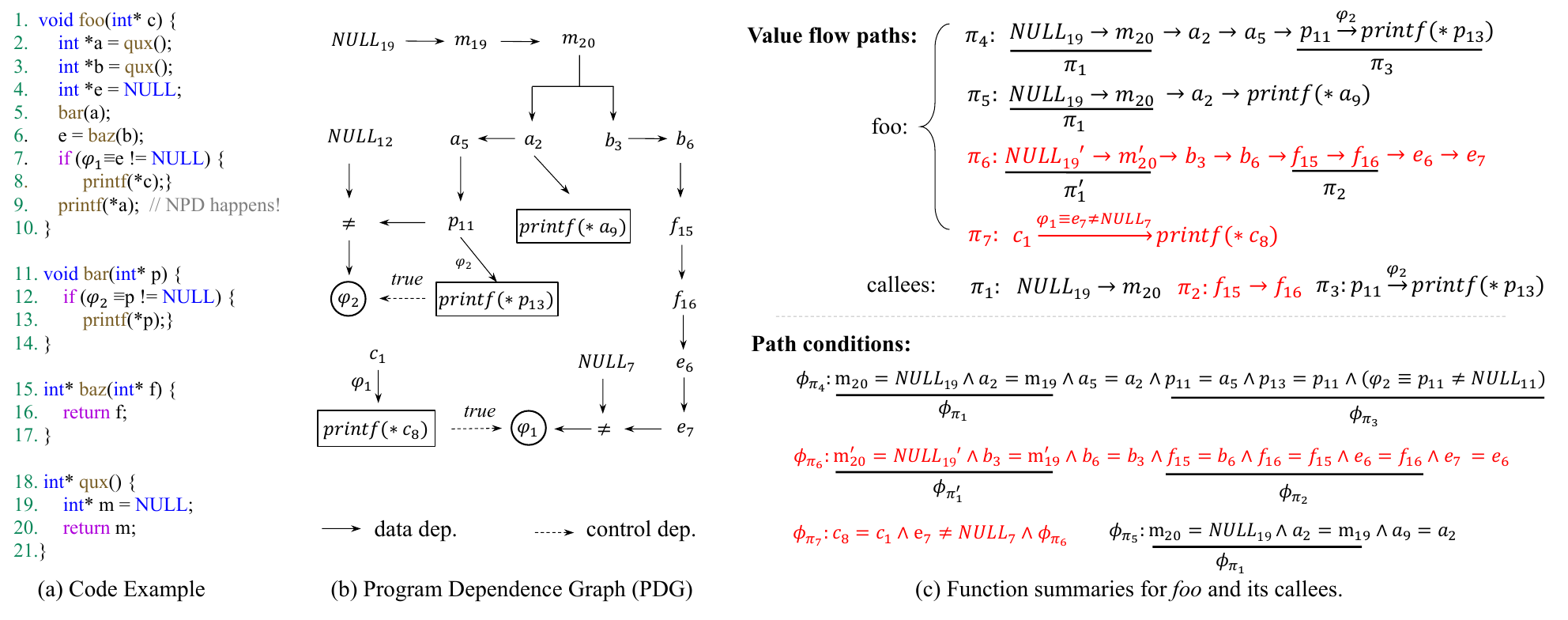}
    \caption{Bottom-up analysis for the code shown in (a).
    The (b) shows the corresponding program dependence graph (PDG). 
    (c) shows the partial function summaries collected during the bottom-up analysis.
    Redundant summaries are highlighted in red.}
    \label{fig:motivation_ex}
\end{figure*}

We use the buggy program shown in Fig.~\ref{fig:motivation_ex}(a) to illustrate the existing bottom-up compositional value flow analysis.
We use the symbol $\pi$ to represent a value-flow path, while $\phi$ and $\varphi$ represent path conditions. Moreover, the variable $v_{i}$ denotes that variable $v$ is either used or defined at Line $i$.
Specifically, one of the function summaries for \textit{foo}, represented as $\pi_{4}$ in Fig.~\ref{fig:motivation_ex}(c), summarizes the propagation path of variable $a_{2}$. 
The variable $a_{2}$ receives the return value from the \textit{qux} function in Line 2 and is subsequently passed to the \textit{bar} function in Line 5.
On one hand, the summary path $\pi_{4}$ is generated by combining the summaries $\pi_{1}$ and $\pi_{3}$, which are collected during the analysis of the \textit{qux} and \textit{bar} functions, respectively, before analyzing the \textit{foo} function.
On the other hand, the summary condition $\phi_{\pi_{4}}$ is obtained by instantiating edges and the corresponding guards along the summary path $\pi_{4}$~\cite{shi2018pinpoint}. 
The guard $\varphi_2$ of the edge $p_{11} \rightarrow printf(*p_{13})$ is the constraint of $p_{11} \neq NULL_{12}$, which is instantiated when collecing the summary $\pi_{3}$ by traversing from vertex $\varphi_2$ on the program dependence graph, shown in Fig.~\ref{fig:motivation_ex}(b).
Once the summary condition $\phi_{\pi_{4}}$ is instantiated, the summary is verified by the constraint solver \textit{Z3}~\cite{de2008z3} before it is stored.
The summary conditions $\phi_{\pi_{1}}$, $\phi_{\pi_{2}}$, and $\phi_{\pi_{3}}$ are also verified when collected.
If a summary condition is unsatisfiable(\textit{unsat}), it is discarded to avoid an unfeasible summary being maintained.
(We provide such a case in Appendix Section~\ref{app:unsat-summary}.)
Moreover, the summary $\pi_1$ of the \textit{qux} function is cloned twice (cloned one denoted as $\pi_1^{\prime}$) with different calling contexts (Line 2 and Line 3) to account for the different propagation paths between the summaries $\pi_4$ and $\pi_6$ (which summarizes the propagation path of variable $b_{3}$). 
This cloning mechanism eliminates the necessity of searching for and analyzing the \textit{qux} function again, leading to improved efficiency.

\textbf{The Explosive Summary Problem.}
However, the bottom-up approach still faces challenges in terms of analysis time and memory consumption. 
According to a report~\cite{shi2018pinpoint}, a single analysis for a project with millions of lines of code can take several hours and require hundreds of gigabytes (GB) of memory. 
The main reason for this is the exponential growth in the size of summaries due to cloning for different calling contexts.
Even worse, exploded summaries lead to frequent calls to the constraint solver, as each summary collection necessitates a call to the solver.
For example, in the previous example, the summary $\pi_1$ is cloned and stored as two copies ($\pi_1$ and $\pi^{\prime}_{1}$) within the function \textit{foo}.
This leads to the collection of three new summaries $\pi_4$, $\pi_5$, and $\pi_6$, which eventually result in three calls to the constraint solver.
When analyzing higher-level callers, the need for additional clones can cause significant performance issues.

Existing techniques to improve the performance have focused on achieving efficient summary path collection~\cite{shi2020pipelining, tang2023Scaling} and the verification of summary conditions \cite{shi2021path}.
Specifically, Shi~\cite{shi2020pipelining} and Tang~\cite{tang2023Scaling} proposed a parallel algorithm to accelerate summary path collection, reducing the analysis time. 
Shi's recent work~\cite{shi2021path} introduced a unified representation of summary paths and summary conditions on the program dependence graph, enabling the direct verification of summary conditions on the PDG and eliminating the need for additional computation and storage of summary conditions.

\textbf{Our Approach.}
To tackle this problem, we propose the first approach that identifies and eliminates ``useless'' summaries while also reducing the size of collected summaries from callee functions.
Our key observation is that certain summaries in the callee functions do not contribute to any source-sink path (even when their summary condition is satisfiable(\textit{sat})) and, thus, can be safely ignored without compromising the analysis's precision.
For example, in Fig.~\ref{fig:motivation_ex}(c), the summary $\pi_6$ obtained from the function \textit{foo} does not lead to any bugs since there are no dereference operations on the inlined $NULL_{19}^{'}$ from the callee function $qux$. 
As a result, it is unnecessary to compute the summary $\pi_6$ and solve its path condition $\phi_{\pi_6}$.
Additionally, we can further reduce unnecessary computations by avoiding the cloning of $\pi_1$, not collecting $\pi_2$ (and solving the summary condition) to eliminate the redundant $\pi_6$.
Our experiments (as shown in Table~\ref{table:benchmark} under the ``\#Redun'' column) indicate that approximately 20\% of redundant summaries are computed, solved, and maintained throughout the analysis process on average.

The benefit of our approach is twofold. 
First, our approach efficiently reduces time and memory usage by eliminating unnecessary computations of summaries, which can become exponentially large as the analysis progresses.
Second, the high-cost invocation of a constraint solver to verify ``useless'' summaries is subsequently avoided.
These two unique advantages make our approach more practical for efficiently analyzing large-scale software systems.
Our approach is orthogonal and can be used in conjunction with other approaches, 
such as enhancing the summary representation 
by employing advanced data structures, i.e., graph structures, for representing and resolving summary conditions~\cite{shi2021path}, and effectively collecting summaries
in a parallel~\cite{tang2023Scaling} or pipelined manner~\cite{shi2020pipelining}. 




\textbf{Challenges and Solutions.} 
The challenge lies in identifying redundant summaries precisely and efficiently without hurting the precision of the analysis.
Specifically, determining the contribution of a summary often relies on information from upper-layer functions that have not been analyzed yet. 

Our key insight is that a useful summary should \emph{be a component of} paths or path conditions associated with a source-sink path of interest.
Specifically, the summary should be reachable from at least one pair of sources and sinks or derived from the path conditions of the source-sink paths.
\sloppy
In Fig.~\ref{fig:motivation_ex}, the usability of summary $\pi_2$ is decided by evaluating its reachability with the source-sink pairs $(NULL_{19}, printf(*a_9))$ and $(NULL_{19}, printf(*p_{13}))$. 
That is, we assess the reachability of the source $NULL_{19}$ and the sink $printf(*a_9)$ (or $printf(*p_{13})$) using the head $f_{15}$ and tail $f_{16}$ of $\pi_2$, which are parameter and return of function $baz$. 
They are represented as $(NULL_{19}, f_{15})$, $(f_{16}, printf(*a_9))$, and $(f_{16}, printf(*p_{13}))$.
Without considering how $\pi_2$ will be used in caller \textit{foo}, we can still decide that $\pi_2$ is not reachable from the two mentioned source-sink pairs. 
Consequently, $\pi_2$ is deemed redundant in the long run and can be promptly discarded.
Using this insight, we have devised a principled, sound, and efficient contribution identification algorithm powered by a novel concept, namely \textit{contribution abstraction}, to identify the contributing summaries.
We give more details in Section~\ref{sec:overview}.

\textbf{Results.}
We have implemented the contribution identification (CI) algorithm based on the state-of-the-art value flow analysis, Fusion~\cite{shi2021path}, and evaluated it on 17 real-world programs.
The evaluation results show our CI algorithm can significantly reduce the time and memory overhead of the Fusion by 45\% and 27\%, respectively. 
Furthermore, the CI algorithm can efficiently identify almost 80\% of redundant summaries while only incurring a minimal additional overhead.
In the largest project, the \textit{mysqld} case, CI helps Fusion save 8107 seconds (2.25 hours) with only 17.31 seconds of overhead, resulting in a ratio of time savings to paid overhead (performance gain) of 468.48 $\times$. Overall, CI achieves a substantial average performance gain of 632.1 $\times$.

To sum up, this paper makes three main contributions:
\begin{itemize}
    \item We identify and address the redundant summary deficiency in the prior value flow analysis.
    \item We design the contribution identification algorithm to identify redundant summaries efficiently and effectively.
    \item On average, the contribution identification algorithm can substantially enhance the performance of value flow analysis, reducing time consumption by 45\% and minimizing memory utilization by 27\%.
\end{itemize}

\section{Background and Preliminary}
\label{sec:problem-formu}

This section introduces the background of path-sensitive value flow analysis and basic notations throughout the paper.
\vspace{-0.7em}
\subsection{Background}
We assume that the target program is in the static single assignment (SSA) form~\cite{cytron1991efficiently}, where each variable has only one definition and multiple definitions are merged using a $\phi$-assignment, following many existing works~\cite{babic2008calysto, cadar2008klee, shi2018pinpoint, sui2016svf, sui2014detecting, xie2005scalable}.
All elements within an array or a union structure are considered to be aliases.
In our implementation,  we have utilized the existing methods to resolve points-to relations~\cite{shi2018pinpoint}.

\textbf{Program Dependence Graph (PDG).}
Given the program $P$, PDG is constructed to characterize how a value flows from one program statement to another through edges labeled with path constraints.
We follow the previous works~\cite{sui2016svf, sui2014detecting, shi2018pinpoint} to construct the program dependence graph, where the definition, the use of all variables, and operators are modeled as vertices.



\begin{definition}[]
\label{def:pdg}
A program dependence graph for a function is a directed graph denoted as $G=(V, O, E_d, E_c)$:
\begin{itemize}
    \item $V$ is a set of vertices, each of which is denoted by $v@s_i$, meaning the variable $v$ is defined or used at a statement $s_i$.
    We write $v@s_i$ as $v_i$ for short as the program is in the SSA form.
    The guard vertices are denoted as $V_{\text{g}} \subseteq V$.
    \item $O$ is a set of operator vertices (binary $\oplus$ or unary $\otimes$), each of which represents a symbolic expression.
    \item $E_d \subseteq (V \cup O) \times (V \cup O)$ is a set of directed edges.
    $(v_i, v_j) \in E_d$ means that the value $v_i$ flows to value $v_j$.
    Edges are labeled with guards from $V_{\text{g}} \cup true $, which represent the constraints that qualify the value flow.
    \item $E_c \subseteq V \times V_b$ is a set of control dependence edges.
\end{itemize}

\end{definition}


\begin{example}
    In Fig.~\ref{fig:motivation_ex}(b), the two value flows from $c_1$ to $printf(*c_8)$ and from $p_{11}$ to $printf(*p_{13})$ are qualified by the branch expressions at Line 7 and Line 12. 
    Therefore, the two edges $(c_1, print(*c_8))$ and $(p_{11}, printf(*p_{13}))$ are labeled by guards $\varphi_1$ and $\varphi_2$. 
    The expression of these branches can be derived by searching the PDG from guard vertices $\varphi_1$ and $\varphi_2$, which are $e_7 \neq NULL_{7}$ and $p_{11} \neq NULL_{12}$, respectively.
    \label{example:pdg}
\end{example}

To check a given value-flow path $\pi$~\cite{shi2018pinpoint}, the path condition $\phi_{\pi}$ incorporates not only the value flows represented by the edges but also the value flows related to the instantiation of the guard vertices labeled on those edges. 
Note that these two categories of value flows are not identical. The value flows related to the instantiation of guard vertices can be interconnected with the value flows of other paths. Thus, the path condition $\phi_{\pi}$ often relates more paths beyond just $\pi$.



\begin{example}
\label{example:path_condition}
    Recall Fig.~\ref{fig:motivation_ex}(c),
    the summary $\pi_7: c_1 \xrightarrow{\varphi_1} printf(*c_8)$ summaries the value flow from $c_1$ to $c_8$ with the guard $\varphi_1$. 
    The value flow represented by the edge is encoded as $c_8 = c_1 \land \varphi_1$.
    When instantiating the constraint represented by the guard $\varphi_1$, which states that $e_7 \neq NULL_{7}$, the value flow of $e_7$ is tracked.
    This results in the identification of the summary path of $\pi_6$, represented as $NULL^{\prime}_{19} \rightarrow m^{\prime}_{20} \rightarrow \dots \rightarrow e_{6} \rightarrow e_7$. 
    Consequently, the path condition of $\pi_6$ is applied in this context. 
    Taking all of this into consideration, the path condition of $\pi_7$ can be expressed as $c_8 = c_1 \land e_7 \neq NULL_{7} \land \phi_{\pi_{6}}$.
    Summary condition $\phi_{\pi_{7}}$ involves more complex value flow than summary path $\pi_{7}$ due to instantiating the constraint represented by the guard $\varphi_1$.
\end{example}

Given a value-flow path $\pi$ on the program dependence graph $G$, $\pi[i]$ represents $i-$th vertex $v_i@s_i$ on the path. 
Specifically, we use $\pi[-1]$ to denote the tail element of $\pi$.
Given sets $V_1$ and $V_2$, which are subsets of the vertices $V$ in the PDG, we use $\Pi(V_1, V_2)$ to represent the set of value-flow paths from a vertex in $V_1$ to another vertex in $V_2$.

The path conditions of a set of value-flow paths $\Pi(V_1, V_2)$ are represented as $\Phi(V_1, V_2, V_{\text{g}})$, 
The additional $V_{\text{g}}$ denotes the set of guard vertices that necessitate instantiation during the construction of the path conditions. 


\textbf{Bottom Up Value Flow Analysis.}
Given the PDG and bug-specific sources $V_{\text{src}}$ and sinks $V_{\text{sink}}$, the path-sensitive value flow analysis is to collect $\Pi(V_{\text{src}}, V_{\text{sink}})$ and $\Phi(V_{\text{src}}, V_{\text{sink}}, V_{\text{g}})$.
To determine the presence of bugs, each path $\pi$ in $\Pi(V_{\text{src}}, V_{\text{sink}})$ is evaluated by checking its path condition $\phi_{\pi}$ using a constraint solver.
If the path condition $\phi_{\pi}$ is determined to be \textit{sat}, the path $\pi$ is reported as a detected bug.
To scale up the collection of $\Pi(V_{\text{src}}, V_{\text{sink}})$ and $\Phi(V_{\text{src}}, V_{\text{sink}}, V_{\text{g}})$, the existing path-sensitive approaches~\cite{cousot2002modular, xie2005scalable, shi2018pinpoint, shi2020conquering, shi2020pipelining, shi2021path} use a compositional manner that analyses each function on a call graph from the bottom.
Note that existing bottom-up analyses~\cite{cousot2002modular, xie2005scalable, shi2018pinpoint, shi2020conquering, shi2020pipelining} first compute the Strongly Connected Components (SCC) of the call graph to make it acyclic.
Then, the path-sensitive methods compute symbolic summaries for each function. 
These summaries are subsequently instantiated in the callers' different contexts, allowing them to be reused to merge various source-sink paths and their corresponding path conditions, thereby eliminating the redundant re-analysis of each function.

However, the alternative top-down approaches~\cite{reps1995precise, reps1994speeding, murphy1999program, sagiv1996precise} analyze functions in a call graph from top to bottom, producing summaries for specific program contexts that cannot be reused for all source-sink paths. Thus, they require analyzing the same function multiple times for different calling contexts, sacrificing path sensitivity. 
As a result, these approaches can only determine the reachability of a source-sink pair without providing the connecting paths and path conditions.

\begin{algorithm}[t]\small
  \caption{Path-Sensitive Value Flow Analysis}
  \label{alg:ps-vf-analysis}
  \SetKwFunction{Q}{pathSensitveAnalysis}
  \SetKwProg{Proc}{Procedure}{}{}
  \SetKwComment{Comment}{//}{}
  \Proc{\Q{P}} {
    Build call graph CG and PDG of {P}\;
      $S(V_{\text{src}}, V_{\text{sink}}) \leftarrow \emptyset$\;
      \ForEach{$f \in CG$}{
      \resizebox{0.8\columnwidth}{!} {
            $S^{f}(V_{\text{fp}}, V_{\text{fr}}) \leftarrow \emptyset$;
            $S^{f}(V_{\text{fp}}, V_{\text{sink}}) \leftarrow \emptyset$;
            $S^f(V_{\text{src}}, V_{\text{fr}})$\;}
      }

    \ForEach{$f$ $\in$ $CG$ in bottom-up order} {
        \ForEach{c $\in$ callees of $f$}{
        \texttt{collectCloneSolveSmry}($S^f(V_{\text{fp}}, V_{\text{fr}}),c$)\;
        \texttt{collectCloneSolveSmry}($S^f(V_{\text{fp}}, V_{\text{sink}}),c$)\;
        \texttt{collectCloneSolveSmry}($S^f(V_{\text{src}}, V_{\text{fr}}),c$)\;
        \texttt{collectSrcSinkPath}($S(V_{\text{src}}, V_{\text{sink}}), c$)\;
        }
    }
    $ \forall (\pi, \varphi) \in S(V_{\text{src}}, V_{\text{sink}}),\, \text{report}\, \pi\, \text{as}\, \text{a}\, \text{bug}\, \text{if}\, \varphi\, \text{is}\, \textit{sat;} $
  }
\end{algorithm}

To construct complete source-sink paths, bottom-up approaches gather three types of summaries.
\begin{definition}[Function Summary]
A summary for the function $f$ is represented by a tuple $s = (\pi, \phi_{\pi})$, 
where summary path $\pi$ captures a value flow path after the callee functions' summaries are cloned.
The summary condition $\phi_{\pi}$ encodes value flows of $\pi$ and value flows instantiated from guard vertices that are labeled on $\pi$.
\begin{itemize}
    \item Transfer summary $S(V_{\text{fp}}, V_{\text{fr}})$ summarizes value flow paths from the function's formal parameters $V_{\text{fp}}$ to the function's formal return $V_{\text{fr}}$. 
    \item Input summary $S(V_{\text{fp}}, V_{\text{sink}})$ summarizes value flow paths from the function's formal parameters to a sink within the function or its callee functions.
    \item Output summary $S(V_{\text{src}}, V_{\text{fr}})$ summarizes value flow paths from sources to the function's formal return. The sources are found within the function or callee functions.
\end{itemize}
\end{definition}

Existing work~\cite{shi2020conquering,shi2018pinpoint} shows that collecting these different categories of value flow paths is sound for bug detection.
Corresponding to $V_{\text{fp}}$ and ${V}_{\text{fr}}$ that represent the sets of formal parameters and formal return,
${V}_{\text{ap}}$ and ${V}_{\text{ar}}$ represent the sets of actual parameters and actual return.
We denote total summaries that are collected from function $f$ as $S^{f}(V_{h}, V_{t})=(\Pi^{f}, \Phi^{f})$.
$V_{h}$ and $V_{t}$ are the head vertices and tail vertices of the summary path that could be collected in Algorithm~\ref{alg:ps-vf-analysis}.
Specifically, $V_{h}$ are the head vertices come from $V_{\text{fp}}$, $V_{\text{ar}}$ and $V_{\text{src}}$, and $V_{t}$ are the tail vertices come from $V_{\text{ap}}, V_{\text{fr}}, V_{\text{sink}}$, and $V_{\text{g}}$.

Algorithm~\ref{alg:ps-vf-analysis} presents the existing bottom-up summary collection~\cite{cousot2002modular, xie2005scalable, shi2018pinpoint, shi2020conquering, shi2020pipelining, shi2021path} of $ S(V_{\text{src}}, V_{\text{sink}})$ for the given program $P$. 
In general, it is accomplished by two helper functions: \texttt{collectCloneSolveSmry} and \texttt{collectSrcSinkPath}. 
The first function, \texttt{collectCloneSolveSmry}, is responsible for collecting summaries by cloning the callee's summaries and then solving the summary condition, filtering out the \textit{unsat} ones.
The second function, \texttt{collectSrcSinkPath}, collects the source-sink paths that can be discovered after collecting the summaries in the current function.
The overall Algorithm~\ref{alg:ps-vf-analysis} begins by constructing the CG and PDG for the program $P$. 
It then initializes a global set $S(V_{\text{src}}, V_{\text{sink}})$ to maintain all the source-sink paths, as well as three summary sets for each function to maintain the three types of summaries. 
The algorithm proceeds to process each function in a bottom-up fashion, collecting three types of function summaries (in Lines 8, 9, and 10) and source-sink paths (in Line 11), assisted by the two helper functions.
Finally, the algorithm reports bugs in Line 12 by solving the path condition after all functions have been analyzed.

The \texttt{collectCloneSolveSmry} helper collects summaries directly from the current function if the value flow path does not pass through a function call. Otherwise, it collects summaries by concatenating the inlined summaries from the called function $c$. 
For example, in Fig.~\ref{fig:motivation_ex}, the summaries $\pi_{1}$ $\pi_{2}$, and $\pi_{3}$ are collected directly from their respective functions, while others are collected by concatenating callee summaries.

A single summary is collected in two steps: collecting the summary path $\Pi$ and instantiating the summary condition $\Phi$. 
Instantiating the summary condition often requires additional summary paths. 
As demonstrated in Example~\ref{example:path_condition}, instantiating a summary condition often involves additional value flow starting from the guard vertex labeled on the summary path. 
Once a condition is instantiated, it is solved by a constraint solver. 
If a summary's condition is \textit{unsat}, the summary is discarded because an infeasible summary implies that the resulting source-sink path is also infeasible.
The helper function puts the feasible summaries into three types of summary sets, which would be used in the caller functions in the incoming analysis.
At that time, the summaries are inlined, which helps the caller form long summaries and possible source-sink paths.

\begin{example}
Recall Fig.~\ref{fig:motivation_ex}(c). 
Functions \textit{qux}, \textit{baz}, and \textit{bar} are called by the function \textit{foo}. 
Thus, \textit{qux}, \textit{baz}, and \textit{bar} are analyzed first, while \textit{foo} is analyzed later.
In function \textit{qux}, only the output summary $s_{1}$ is generated between the source $V_{\texttt{src}} = \{NULL_{19}\}$ and the formal return $V_{\texttt{fr}} = \{m_{20}\}$, and then $\phi_{\pi_{1}}$ is verified.
Thus, \textit{qux} has $S^{\texttt{qux}}(V_{\texttt{src}}, V_{\texttt{fr}}) = \{s_1=(\pi_{1},\phi_{\pi_{1}})\}$, with the other three sets being empty.
In function \textit{baz}, only the transfer summary $s_{2}$ is collected between the formal parameter $V_{\texttt{fp}} = \{f_{15}\}$ and the formal return $V_{\texttt{fr}} = \{f_{16}\}$, with the condition verified.
Thus, \textit{baz} has $S^{\texttt{baz}}(V_{\texttt{fp}}, V_{\texttt{fr}}) = \{s_2=(\pi_{2},\phi_{\pi_{2}})\}$, with the other three sets being empty.
In function \textit{bar}, the input summary $s_{3}$ is collected between the formal parameter $V_{\texttt{fp}} = \{p_{11}\}$ and the sink $V_{\texttt{sink}} = \{printf(*a_{9})\}$, with condition verified.
Thus, \textit{bar} owns $S^{\texttt{bar}}(V_{\texttt{fp}}, V_{\texttt{sink}}) = \{s_3=(\pi_{3}, \phi_{\pi_{3}})\}$, with the other three sets being empty.
Since there is a conditional edge in $\pi_{3}$, its summary condition $\phi_{\pi_{3}}$ involves the instantiation of the constraint represented by the guard $\varphi_2$, which is $p_{11} \neq NULL_{11}$. This involves the value flow path of $p_{11}$, which coincides with its summary path.

\end{example}

\begin{example}
    When analyzing the top layer function \textit{foo}, the summaries are cloned from the bottom layer functions accordingly.
    When collecting the transfer summaries starting from formal parameter $V_{\texttt{fp}} = \{ c_1 \}$, which reaches a sink $printf(*c_8)$ but cannot reach any actual input in $V_{ap} = \{ a_{5}, b_{6} \}$, no transfer summaries are gathered as a result.

    When collecting the input summaries, the summary $s_{7}$ is generated with a summary path collected from the formal parameter $V_{\texttt{fp}} = \{c_1\}$ to the sink $V_{\texttt{sink}} = \{printf(*c_8)\}$, without cloning callee summaries. 
    However, the summary condition $\phi_{\pi_7}$ necessitates instantiating the guard vertex $\varphi_{1}$, where $V_{g} = \{\varphi_{1}\}$.
    To this end, the value flow starting from $\varphi_{1}$ is tracked. 
    When reaching the actual output $e_6$ of the function call to \textit{bar} in line 6, the summary $s_{2}$ is cloned from the callee \textit{bar}. 
    Moving forward, when reaching the actual output $b_3$ of the function call to \textit{qux} in line 3, the summary $s_{1}$ is cloned, thus forming the summary $s_{6}$.
    Finanly $\phi_{\pi_7}$ is verified.

    When collecting the output summaries, because the $V_{\texttt{src}}$ in the current function is empty, the output summaries of the callee function are cloned, introducing additional sources. 
    Thus, in line 2, the summary $s_{1}$ is inlined again, which is the output summary of the function \textit{qux}.
    By combining the summary $s_{1}$, the value path from the output $a_2$ to $a_5$ is traced and involves the call to the function \textit{bar} in line 5. 
    Since $a_5$ is passed to \textit{bar}, the input summary or transfer summary that starts at the corresponding formal parameter is inlined. 
    In our case, the input summary $s_{3}$ of the function \textit{bar} is inlined. 
    After concatenating with the input summary, the path reaches a sink, forming a complete source-sink path, but does not reach any formal return. Thus, no output summary is collected.
    
\end{example}

The \texttt{collectSrcSinkPath} helper is similar to the \texttt{collectCloneSolveSmry} but tries to collect the source-sink paths in each iteration and does not solve the path conditions. 
In the example of Fig.~\ref{fig:motivation_ex}, the helper collects no source-sink paths from bottom-layer functions, as no such paths are formed, but it collects two source-sink paths, $s_4$ and $s_5$, from the upper-layer function \textit{foo}.
Eventually, $\phi_{\pi_4}$ and $\phi_{\pi_5}$ are solved when reporting bugs in Line 12 of Algorithm~\ref{alg:ps-vf-analysis}.


\section{Overview}
\label{sec:overview}
In this section, we illustrate the problem using the motivating example and briefly describe our key idea.

\subsection{Explosive Summary Problem}
To detect the bug, the bottom-up collection of function summaries outlined in Algorithm~\ref{alg:ps-vf-analysis} can collect and maintain a superset of the function summaries that are actually required.
As highlighted in red in Fig.~\ref{fig:motivation_ex}(c), the summaries $s_2$, $s_6$, and $s_7$ do not contribute to the two source-sink paths, $s_{4}$ and $s_{5}$, for detecting the NPD bug, i.e., either as components of these paths or their associated path conditions.
Specifically, non-contributing summaries can arise in two scenarios in Algorithm~\ref{alg:ps-vf-analysis}.
First, when collecting three types of summaries, over-summarization can occur.
For instance,
when analyzing the function $baz$, there is no prior knowledge of which the specific summary contributes, resulting in the conservative collection of all summaries within it.
Consequently, the non-contributing summary $s_2$ is collected as a transfer summary.
Second, non-contributing summaries can be induced through the cloning of callee summaries. 
For example, the summary $s_6$ is a non-contributing summary generated by cloning and concatenating with the callee summary $s_2$. 
Additionally, given that there are no source-sink paths within the function $boo$ that rely on $s_2$, and considering that $s_6$ is deemed non-contributory, the cloning of $s_2$ from function $baz$ should be avoided during the analysis of \textit{foo}.
As summaries are maintained and cloned into higher-level functions, the size of $S^f$ can exponentially increase due to the explosion of paths.
More importantly, collecting non-contributing summaries introduces expensive constraint-solving.
To sum up, collecting and cloning only the contributing summaries can significantly improve scalability.
The challenge lies in identifying redundant summaries precisely and efficiently without hurting the precision of the analysis.

\subsection{Removing Redundant Summaries}

We first propose the key idea of assessing whether a summary is contributing or not by solving two graph reachability problems between heads and tails of the summary $s$ with distinct reaching targets (source-sink pairs or guards) without computing any summaries.
Based on the graph reachability abstraction, we design the contribution identification algorithm, which identifies a set of necessary head and tail vertices for identifying the contributing summaries. 
Summaries that are collected outside these necessary head and tail vertices are identified as non-contributing automatically.
Next, we explain how our graph reachability abstraction and algorithmic design overcome the above challenges.

\textbf{Contributing Summmary.}
Our key observation is that whether a summary is contributing hinges on two aspects: its path contribution, where it must be a component of source-sink paths, and its condition contribution, where it must be involved in the path conditions associated with a source-sink path.
Thus, we establish the definition of a contributing summary generated from a function based on its contribution to the source-sink paths $S(V_{\text{src}}, V_{\text{sink}})$.

\begin{definition}[Contributing Summary] \label{def:contributing_summary}
A summary $s \in S^f$ is considered a contributing summary for function $f$ if it satisfies at least one of the following criteria:
(1) Path Contribution: The summary path $\pi$ is used to connect at least one source-sink path, i.e., $\pi \in \Pi(V_{\text{src}}, V_{\text{sink}})$.
(2) Condition Contribution: The summary path $\pi$ is used to instantiate at least one guard vertex labeled along the source-sink paths, i.e., $\phi_{\pi} \in \Phi(V_{\text{src}}, V_{\text{sink}}, V_{\text{g}})$.
\end{definition}

We use Fig.~\ref{fig:motivation_ex} as an example.
The path $\pi_{1}$ is reachable from both source-sink pairs $(NULL_{19}, printf(*p_{13}))$ and $(NULL_{19}, printf(*a_9))$.
In addition, the path $\pi_3$ is reachable from the source-sink pair $(NULL_{19}, printf(*p_{13}))$.
Therefore, they are identified as contributing summaries; indeed, they are components of two source-sink paths, $\pi_4$ and $\pi_5$.
Comparatively, the paths $\pi_{2}$, $\pi_{6}$, and $\pi_{7}$ are neither reachable by any pair of sources and sinks nor reachable by the guard vertex $\varphi_2$ labeled on the source-sink path $\pi_4$. 
Thus, these paths $\pi_{2}$, $\pi_{6}$, and $\pi_{7}$ are identified as non-contributing summaries.
Additionally, despite the reachability of the summary $\pi_{6}$ from the guard vertex $\varphi_1$ labeled on $\pi_7$, more specifically, the tail of $\pi_{6}$, denoted as $(e_7)$, being reachable from $\varphi_1$, the summary $\pi_{7}$ does not contribute to any source-sink path or conditions. 
Consequently, it becomes redundant for NPD detection.

In summary, the summary contribution is identified by assessing two reachabilities between specific heads and tails of the summary path with various targets:
\begin{enumerate}
    \item Path contribution: If the summary $s=(\pi, \phi_{\pi})$ has the path contribution, a source-sink pair ($(src, sink) \in V_{\text{src}} \times V_{\text{sink}}$) exists, where $src$ can reach both $\pi[0]$ and $\pi[-1]$, and $sink$ is reached by both $\pi[0]$ and $\pi[-1]$.
   \item Condition contribution: If the summary $s=(\pi, \phi_{\pi})$ has the condition contribution criteria, there exists a guard vertex $\text{g} \in V_{\text{g}}$ from $\Phi(V_{\text{src}}, V_{\text{sink}}, V_{\text{g}})$ that are reached by $\pi[0]$ and $\pi[-1]$.
\end{enumerate}

With the above abstraction, the assessment of contributing summaries is reduced to two graph reachability problems. 
In Section~\ref{sec:contribution-indentify}, we give a sound, efficient, and effective contribution identification algorithm by applying the abstraction.

\section{Contribution Identification}
\label{sec:contribution-indentify}

In this section, we first present three technical designs of our contribution identification algorithm.
We then give the details of the identification algorithm that identifies the necessary vertices for path and condition contribution.
Finally, we establish the soundness of our approach, analyze the complexity of algorithms, and discuss the advanced graph reachability with consideration of the calling context.

\textit{\textbf{Preserving the precision of the analysis.}} To ensure this, instead of directly utilizing the abstractions to identify non-contributing summaries,
    the identification algorithm uses the abstractions to soundly identify all necessary head vertices $V_{h}$ and tail vertices $V_{t}$ that are reached by source-sink pairs (path contribution) as well as guard vertices $V_{g}$ that are labeled on source-sink paths (condition contribution).
    This means that contributing summaries can only be collected within necessary vertices, which we denote as $V^{N}$.
    Therefore, summaries contributing to source-sink paths can be collected within $V^{N}$ as in the traditional methods.
    In contrast, summaries collected outside $V^{N}$ are considered as the non-contributing summaries.
    The soundness proof is given in Section~\ref{sec:sound}.

\textit{\textbf{Efficient and effective identification.}}
    The identification process relies on graph reachability. 
    More precise graph reachability results in fewer necessary vertices $V^{N}$ being identified, allowing for recognizing more non-contributing summaries starting outside of $V^{N}$. 
    However, using advanced reachability algorithms increases the identification overhead. 
    The complexity of more advanced reachability algorithms often outweighs the precision gains they can provide. 
    To strike a balance, i.e., spending minimal overhead while significantly boosting the efficiency of path-sensitive analysis, we select the classic breadth-first search (\texttt{bfs}) algorithm for implementing our abstractions. More discussion about this is in Section~\ref{sec:discussion}.

\textit{\textbf{Resolving implicit contribution.}}
    The source of the implicit contribution comes from the condition contribution of a summary.
    With the abstractions, resolving the implicit contribution is transferred to gather the necessary guard vertices that are labeled on source-sink paths.
    The key is that the necessary guard vertices could be obtained from edge sets that are reachable from the necessary heads and tails for path contribution.

The contribution identification algorithm is outlined in Algorithm~\ref{alg:main}.
At a high level, it identifies necessary vertices $V^{N}$ in two parts for path and condition contribution, respectively, through three stages.
First, it identifies the first part of necessary vertices $V^{N}$ for path contribution using the procedure \texttt{identifyPathContrib} in Algorithm~\ref{alg:main}.
Next, using these necessary heads and tails for path contribution, we collect the necessary guard vertices with the procedure \texttt{collectNecGuards} in Algorithm~\ref{alg:condition}.
Lastly, based on the necessary guard vertices and heads and tails that are not identified for path contribution, we further identify the second part of necessary vertices for condition contribution using the procedure \texttt{identifyCondContrib} in Algorithm~\ref{alg:condition}.

\subsection{Path Contribution Identification}

\begin{algorithm}[t]\small
  \caption{Contribution Identification}
  \label{alg:main}
  \SetKwData{Vn}{$V^{\texttt{N}}$}
\SetKwData{Vcand}{$V^{\texttt{cand}}$}
  \SetKwFunction{Q}{identifyContrib}
  \SetKwProg{Proc}{Procedure}{}{}
    \Proc{\Q{$G$}} {
    $\Vn \leftarrow \emptyset$; $\Vcand \leftarrow \emptyset$\;
    \texttt{identifyPathContrib}(G,\Vn,\Vcand)\;
    \texttt{identifyCondContrib}(G,\Vn,\Vcand)\;
    \Return $V^{\texttt{N}}$\;
  }

  \SetKwFunction{Q}{identifyPathContrib}
  \SetKwProg{Proc}{Procedure}{}{}
  \Proc{\Q{G,\Vn,\Vcand}} {
    srcVisited $\leftarrow \emptyset $; sinkVisited $\leftarrow \emptyset $\;
    \SetKwData{Vsrc}{$V_{\texttt{src}}$}
    \ForEach{$v \in \Vsrc $} {
        \texttt{bfs}(srcVisited, $v$, G, forward)\;
    }
    \SetKwData{Vsink}{$V_{\texttt{sink}}$}
    \ForEach{$v \in \Vsink $} {
        \texttt{bfs}(sinkVisited, $v$, G, backward)\;
    }

    \Vn $\leftarrow $ {srcVisited} $\cap$ {sinkVisited} $\cap\ (V_{t} \cup V_{h} - V_{g})$\;
    \Vcand $\leftarrow$ {srcVisited} $\cup$ {sinkVisited} $- V^{N}$\;
    
  }
\end{algorithm}



Procedure \texttt{identifyPathContrib} in Algorithm~\ref{alg:main} utilizes \texttt{bfs} to explore the graph separately from both the source and sink vertices. 
Since a vertex $v$ only needs to be reachable by at least one source-sink pair, the \texttt{bfs} initiated from the source (sink) vertices maintain a shared visiting set called $\texttt{srcVisited}$ ($\texttt{sinkVisited}$). 
This ensures that each vertex is visited only once during the \texttt{bfs} from the sources (sinks). 
After completing all \texttt{bfs}, the necessary vertices for path contribution can be obtained by computing the intersection between the vertices visited both by $\texttt{srcVisited}$ and $\texttt{sinkVisited}$ and $V_{t} \cup V_{h} - V_{\text{g}}$ in Line 12.
As the guard vertices could not have the path contribution, the necessary vertices only come from $V_{t} \cup V_{h} - V_{\text{g}}$.
Vertices that are visited by sources and sinks but not identified as necessary vertices for path contribution are called candidates, denoted as $V^{\texttt{cand}}$. 
These vertices may have condition contribution, which are computed on Line 13 and passed to the procedure \texttt{identifyCondContrib} on Line 4.

\subsection{Condition Contribution Identification}
\label{sec:condition-identification}

The necessary guard vertices are labeled on the edges of the source-sink paths. 
Thus, necessary guard vertices can be collected using the necessary vertices for path contribution.

The necessary guards are collected and maintained in $V_{g}^{\texttt{nec}}$ through procedure \texttt{gatherNecGuards}, using the \texttt{bfsEdge} traversal to gather the visited edges. 
Two shared edge sets, \texttt{fwdEdges} and \texttt{bwdEdges}, are utilized to keep track of the visited edges for forward and backward \texttt{bfsEdge} starting from the vertices in $V^{N}$, respectively. 
These two sets ensure that each edge is visited only once during the forward and backward \texttt{bfsEdge}.
For each edge $(u, v)$ encountered in \texttt{fwdEdges}, if the reverse edge $(v, u)$ is also found in \texttt{bwdEdges}, the label $L_d(u, v)$ is added to $V_{g}^{\texttt{nec}}$.


\begin{algorithm}[t]\small
  \caption{Condition Contribution Identification}
  \label{alg:condition}
    \SetKwData{Vn}{$V^{\texttt{N}}$}
    \SetKwData{Vcand}{$V^{\texttt{cand}}$}
    \SetKwData{necG}{$V^{\texttt{nec}}_{g}$}
  \SetKwFunction{Q}{identifyCondContrib}
  \SetKwProg{Proc}{Procedure}{}{}
  \Proc{\Q{G,\Vn,\Vcand}} {
    $\necG \leftarrow \emptyset$; {fwdVisited} $\leftarrow \emptyset$; {bwdVisited} $\leftarrow \emptyset $\;
    \texttt{gatherNecGuards}(\Vn, \necG)\;
    \ForEach{$v$ $\in$ \Vcand} {
        \texttt{bfs}({fwdVisited}, $v$, G, forward)\;
    }
    \ForEach{$v$ $\in$ \necG} {
        \texttt{bfs}({bwdVisited}, $v$, G, backward)\;
    }

    \Vn $\leftarrow$ \Vn $\cup$ ({fwdVisited} $\cap$ {bwdVisited}) $\cap\ (V_{t} \cup V_{h})$\;
    
  }
  
  \SetKwFunction{Q}{gatherNecGuards}
  \SetKwProg{Proc}{Procedure}{}{}
  \Proc{\Q{\Vn, \necG}} {
        {fwdEdges} $\leftarrow \emptyset$; {bwdEdges} $\leftarrow \emptyset$\;
        \ForEach{$v$ $\in$ \Vn} {
            \texttt{bfsEdge}({fwdEdges}, $v$, G, forward)\;
        }
        \ForEach{$v$ $\in$ \Vn} {
            \texttt{bfsEdge}({bwdEdges}, $v$, G, backward)\;
        }
        
        \ForEach{$(u, v) \in \mathrm{bwdEdges}$ $\cap\ \mathrm{fwdEdges}$} {
            \necG $\leftarrow $ \necG $ \cup $ $ L_d(u, v)$\;
        }
  }
\end{algorithm}

After collecting the necessary guard vertices, 
the algorithm then proceeds to collect another part of the necessary vertices for condition contribution.
If a vertex has already been identified for path contribution, separately identifying its condition contribution is unnecessary.
Thus, the procedure \texttt{identifyCondContrib} explores the vertices from the candidates in the forward direction and from each necessary guard vertex in the backward direction.
During the \texttt{bfs} iterations, the shared sets \texttt{fwdVisited} and \texttt{bwdVisited} are utilized to keep track of the visited vertices in each direction, respectively. 
By examining the head and tail vertices present in both \texttt{fwdVisited} and \texttt{bwdVisited} sets, another part of the necessary vertices can be identified.

\begin{example}
Procedure \texttt{identifyPathContrib}'s output is $\{ NULL_{19}, m_{20}, a_2, a_5, \text{printf}(*a_9), p_{11}, \text{printf}(*p_{13}) \}$, and the necessary guard vertice is $\{ \varphi_2 \}$.
The vertices that have condition contribution are $\{ p_{11}, a_2, a_5, m_{20}, NULL_{19}, \varphi_2 \}$ and some of them are included in the path contribution. 
Thus, the output of the procedure \texttt{identifyCondContrib} is $\{ \varphi_2 \}$.
Also, the set $V^{N}$ to assess the contributing summary is $\{ NULL_{19}, m_{20}, \text{printf}(*a_9), p_{11}, \text{printf}(*p_{13}), a_2, a_5, \varphi_2 \}$.
\end{example}

\subsection{Soundness}
\label{sec:sound}

We propose the following theorem to establish the soundness of abstracting contributions in identifying non-contributing summaries for function $f$ based on $V^{N}$.

\begin{theorem}[Soundness]
\label{thm:sound}

Given the set $V^{N}$ identified, for any function $f \in P$, if a summary $s=(\pi, \phi)$ is collected and neither $\pi[0]$ nor $\pi[-1]$ appears in $V^{N}$, it must be a non-contributing summary for function $f$. Canceling the corresponding operations does not affect $S(V_{\text{src}}, V_{\text{sink}})$.

See the proof in Appendix Section~\ref{app:sound}.

\end{theorem}

\begin{example}
In Fig.~\ref{fig:motivation_ex}, collecting summary path $\pi_{6}$ and verifying its condition $\phi_{\pi_{6}}$ are prevented because it is determined that cloning $\pi_{1}^{\prime}$ to concatenate with $b_3 \rightarrow b_6$ would result in a non-contributing summary.
It is because $b_3$ is not included in $V^{N}$.
Similarly, the collection and verification of summaries $s_{2}$ and $s_7$ are removed because the head vertices $f_{15}$ and $c_1$ of these summaries are not present in $V^{N}$.
\end{example}

\begin{algorithm}[t]\small
  \caption{New Path-Sensitive Value Flow Analysis}
  \label{alg:boosted-ps-vf-analysis}
  \SetKwFunction{Q}{newPathSensitveAnalysis}
  \SetKwProg{Proc}{Procedure}{}{}
  \SetKwComment{Comment}{//}{}
  \SetKwInOut{Input}{Input}
  \SetKwInOut{Output}{Output}
  \Proc{ \Q{P} } {
        Build call graph CG and PDG of {P}\;
      $V^{N} \leftarrow $ \texttt{identifyContrib}(PDG)\;
        /* \textbf{\textit{The initialize part is the same as the Algorithm~\ref{alg:ps-vf-analysis}}}  */
        
      \ForEach{$f$ $\in CG$ in bottom-up order} {
            /*~\textit{Pruning the redundancies.} ~*/
            
            $V_{\text{fp}} \leftarrow V_{\text{fp}} \cap V^{N}$; $V_{\text{ap}} \leftarrow V_{\text{ap}} \cap V^{N}$\;
            $V_{\text{fr}} \leftarrow V_{\text{fr}} \cap V^{N}$; $V_{\text{ar}} \leftarrow V_{\text{ar}} \cap V^{N}$; $V_{\text{g}} \leftarrow V_{\text{g}} \cap V^{N}$\;
            /* \textbf{\textit{The summary collection part, which is the same as the Alogirhtm~\ref{alg:ps-vf-analysis}.} } ~*/
      }             
    $ \forall (\pi, \varphi) \in S(V_{\text{src}}, V_{\text{sink}}),\, \text{report}\, \pi\, \text{as}\, \text{a}\, \text{bug}\, \text{if}\, \varphi\, \text{is}\, \text{sat;} $
  }
\end{algorithm}

\subsection{Summary and Discussion}
\label{sec:discussion}


Based on the identification of $V^{N}$, as outlined in Theorem~\ref{thm:sound}, Algorithm~\ref{alg:ps-vf-analysis} can be improved and revised as Algorithm~\ref{alg:boosted-ps-vf-analysis}.
During the process of collecting the three types of summaries, the algorithm incorporates a filtering step for the head, tail, and guard vertices using $V^{N}$. 
The filtering step guarantees that summaries located outside of $V^{N}$ are not collected or cloned, effectively eliminating non-contributing summaries. Verification of these summaries by a constraint solver is thus avoided.
Next, we discuss the important points of the algorithm, including its complexity, precision, and efficiency.


\textbf{Complexity.}
The procedure \texttt{identifyPathContrib} involves searching the PDG $G$ at most twice, as do each of the other procedures, since each edge is visited only once, either in the forward or backward traversal direction. As a result, the overall complexity is linear with respect to the size of PDG.


\textbf{Precision and Efficiency.}
The soundness of $V^{N}$ ensures the correct collection of contributing summaries. 
However, there may still be vertices within $V^{N}$ that lead to non-contributing summaries. 
The precision of $V^{N}$ can be increased by minimizing the number of such vertices, thereby identifying more non-contributing summaries.
The reason for the incorrect collection of some vertices in $V^{N}$ is that Algorithm~\ref{alg:main} utilizes a traditional reachability algorithm, specifically the \texttt{bfs}, to address the reachability problems without differentiating the calling context under which the summary could be used.
However, certain non-contributing summaries can only be identified under specific contexts, i.e., via context-free-language (CFL) reachability~\cite{reps1994speeding, reps1995shape, kodumal2004set, chaudhuri2008subcubic, yannakakis1990graph, melski2000interconvertibility}.
Consequently, non-contributing summaries that are reachable under traditional reachability algorithms but not under context-sensitive reachability algorithms could be further identified. 
The precision of $V^{N}$, therefore, depends on the approach used to solve reachability.
The CFL reachability algorithm takes calling context into consideration by respectively labeling function calls and returns with matched parentheses \([_k\) and \(]_k\) at line $k$.
Thus, a vertex \(i\) is context-sensitively reachable from a vertex \(j\) if the label string of the path does not contain any mismatched parentheses.

\begin{example}
In Fig.~\ref{fig:motivation_ex} (b), the following edges are labeled with parentheses to fulfill the requirements of the CFL reachability algorithm:
$m_{20} \xrightarrow{]_2} a_{2}$, $m_{20} \xrightarrow{]_3} b_{3}$,
$a_{5} \xrightarrow{[_5} p_{11}$, $b_{6} \xrightarrow{[_6} f_{15}$, and $f_{16} \xrightarrow{]_6} e_{6}$.
By replacing \texttt{bfs} with a CFL reachability algorithm in Algorithm~\ref{alg:main}, we would produce $V^{\texttt{N}} = \{NULL_{19}, m_{20}, \text{printf}(*a_9), p_{11}, \text{printf}(*p_{13}), a_2, a_5, \varphi_2 \}$, which is the same as when using \texttt{bfs}. 
In this case, using the CFL reachability algorithm does not improve the precision.
\end{example}

In practice, the precision of $V^{N}$ that can be improved by CFL reachability is limited. In our evaluation, approximately 80\% of non-contributing summaries were successfully identified using \texttt{bfs}.
However, in terms of efficiency, CFL reachability often encounters a cubic complexity barrier~\cite{kodumal2004set, chaudhuri2008subcubic, yannakakis1990graph, melski2000interconvertibility}. 
As demonstrated by our experiment in Fig.~\ref{fig:cfl-light-fusion}, using a CFL-based algorithm to collect $V^{N}$ significantly decreases overall performance. 
Thus, we use \texttt{bfs}.

\section{Evaluation}

We have implemented the contribution identification (CI) algorithm in Algorithm~\ref{alg:main} to identify the path and condition contribution in the state-of-the-art value flow analysis tool Fusion~\cite{shi2021path} for detecting NPD in C/C++ code.
We denote Fusion with CI enabled as Light-Fusion, which serves as the performance-boosted client powered by our technique.
We investigate the following three questions:
\begin{itemize}
    \item (\textbf{RQ1}): How effective and efficient is CI in identifying the summary contribution?
    \item (\textbf{RQ2}): How much can CI boost the performance of existing value-flow analysis?
    \item (\textbf{RQ3}): Can CI enhance the path-sensitive analyzer's performance to be comparable with the top-down approach?
\end{itemize}



\subsection{Experimental Setup}

\begin{table}[t]\small
\caption{
\label{table:benchmark}
$|S^{all}|$ is total size of collected summaries. 
\texttt{\#Redun} is the number of redundant summaries that occur, with the percentage relative to $|S^{all}|$ shown in parentheses.
\texttt{\#Identified} is the number of identified summaries by CI, with the percentage relative to \texttt{\#Redun} shown in parentheses.
\texttt{\#Src} and \texttt{\#Sink} represent the number of sources and sinks.
}
\resizebox{\columnwidth}{!}{
\begin{tabular}{r|lr|rrr|rr}
\hline
ID  & Program    & KLoC & $|S^{all}|$ & \multicolumn{1}{c}{\#Redun}   & \#Identified & \#Src                    & \#Sink \\ \hline
1   & leela      & 21   & 47.2K  & 12.4K(\textbf{26\%})  & 9.3K(\textbf{75\%})  & 15                        & 3.2K    \\
2   & nab        & 24   & 34.6K  & 5.2K(\textbf{15\%})   & 3.7K(\textbf{73\%})  & 88                        & 3.6K    \\
3   & x264       & 96   & 94.2K  & 24K(\textbf{26\%})    & 19K(\textbf{79\%})  & 58                        & 7.7K    \\
4   & wrf        & 130  & 51.1K  & 10.1K(\textbf{20\%})  & 8.4K(\textbf{83\%})  & 123                       & 4.3K    \\
5   & omnetpp    & 134  & 405.6K & 78.1K(\textbf{19\%})  & 66.1K(\textbf{85\%})  & 146                       & 27.5K   \\
6   & povray     & 170  & 231.5K & 45.8K(\textbf{20\%})  & 33.5K(\textbf{73\%})  & 24                        & 14.5K   \\
7   & cactus     & 257  & 1.1M   & 307.7K(\textbf{29\%}) & 257.5K(\textbf{84\%}) & 38                        & 49.8K   \\
8   & imagick    & 259  & 381.4K & 30.3K(\textbf{8\%})  & 22.5K(\textbf{74\%})   & 154                       & 12.2K   \\
9   & perlbmk    & 362  & 1.4M   & 217.8K(\textbf{15\%}) & 197.7K(\textbf{91\%}) & 232                       & 40.9K   \\
10  & cam4       & 407  & 46.7K  & 10.1K(\textbf{22\%})  & 7.2K(\textbf{72\%})  & 53                        & 3.4K    \\
11  & parest     & 427  & 3.6M   & 658.3K(\textbf{18\%}) & 555.9K(\textbf{84\%}) & 62                        & 215.7K  \\
12  & xalanbmk   & 520  & 1.4M   & 294.4K(\textbf{21\%}) & 258.9K(\textbf{88\%})  & 23                        & 77.7K   \\
13  & gcc        & 1304 & 3.5M   & 382.3K(\textbf{11\%}) & 287.4K(\textbf{75\%})  & 181                       & 146.2K  \\
14  & blender    & 1577 & 3.2M   & 605.7K(\textbf{19\%}) & 505.6K(\textbf{83\%})  & 127                       & 182.6K  \\ \hline
15  & libicu     & 537  & 1.1M   & 190.4K(\textbf{17\%}) & 165.8K(\textbf{87\%})  & 307                       & 76.7K   \\
16  & ffmpeg     & 1346 & 2.1M   & 393.6K(\textbf{19\%}) & 244.7K(\textbf{62\%})  & 491                       & 146.3K  \\
17  & mysqld     & 2030 & 3.8M   & 723.6K(\textbf{19\%}) & 519.2K(\textbf{72\%})  & 141                       & 215.4K  \\ \hline
avg &            &      & 1.3M   & 234.7K(\textbf{19\%}) & 186K(\textbf{79\%})  &                           &         \\ \hline
\end{tabular}

}
\end{table}

\textbf{Baselines.}
Fusion collects the summaries in a parallel way, i.e., parallelizing functions located within the same level of the call graph, and uses the graph representation of summary conditions~\cite{shi2021path}.
First, we compare Light-Fusion with Fusion. 
We did not compare to~\cite{shi2020pipelining}\cite{tang2023Scaling} as their methodologies involve generating redundant summaries to enable parallelism, which contradicts the principles and objectives of our approach.
In addition, we have developed a variant of Light-Fusion~(denoted as CFL-Light-Fusion) that uses a more precise identification algorithm achieved through CFL reachability.
Specifically, we adopted the open-source implementation of the state-of-the-art CFL reachability algorithm~\cite{lei2022taming} provided by \cite{pocr} and replaced the \texttt{bfs} used in Algorithm~\ref{alg:main}.
Lastly, we compare the Light-Fusion to \textit{PhASAR}~\cite{10.1007/978-3-030-17465-1_22}, an open-sourced implementation of top-down approaches~\cite{reps1995precise, reps1994speeding, murphy1999program, sagiv1996precise}. 
To mitigate the impact of the execution environment, we ran experiments three times and calculated the average performance and performance gains.

\textbf{Subjects.}
We have included all the large programs from the SPEC CPU@2017 benchmark~\cite{bucek2018spec} that consist of more than 10 KLoC. 
Additionally, we have selected three large programs, namely \textit{libicu}, \textit{ffmpeg}, and \textit{mysqld}, which are widely-used software systems in their respective domains.

\textbf{Sources and Sinks.}
For checking NPD, 
NULL pointers are selected as sources.
The dereference operations are selected as sinks.
Table~\ref{table:benchmark} lists the evaluation subjects with statistics of sources and sinks. 

\textbf{Environment.}
All experiments were run on a server with eighty “Intel Xeon CPU E5-2698 v4@2.20GHz” processors and 512 GB of memory running Ubuntu-18.04.
Each program is analyzed with a limit of 12 hours and 256 GB of memory.
Fifteen threads are used to analyze the functions in the same layer when running both Fusion and Light-Fusion.
The solver used to verify constraints is \textit{Z3}\cite{de2008z3}.


\subsection{RQ1: Effectiveness and Efficiency}

To study the effectiveness of CI, we run two experiments for each benchmark.
In the first experiment, we execute Fusion to produce the results $S(V_{\text{src}}, V_{\text{sink}})$ and collect all potential summaries that could be generated during analysis, denoted as $S^{all}$.
Note that $S^{all}$ contains the non-contributing summaries, which is a superset of $S(V_{\text{src}}, V_{\text{sink}})$.  
The summary size of each benchmark is recorded in the $|S^{all}|$ column of Table ~\ref{table:benchmark}.
Second, we count non-contributing summaries from $S^{all}$ based on the definition of contributing summaries (Definition~\ref{def:contributing_summary}). 
The relative number of non-contributing summaries is reported in the ``\#Redun'' column of Table ~\ref{table:benchmark}.

\begin{table}[]\small
\caption{ 
The running time and memory usage of Fusion (F), Light-Fusion (L-F), and building of PDG are presented, along with CI's running time and performance gains (Gains).}
\label{table:compare}
\resizebox{\columnwidth}{!} {
\begin{tabular}{c|rrr|rrrr|r}
\hline
    & \multicolumn{3}{c|}{Memory(GB)} & \multicolumn{4}{c|}{Time(s)}                       &         \\
ID  & F         & \multicolumn{1}{c}{L-F}       & PDG     & F       & \multicolumn{1}{c}{L-F}     & PDG    & CI & Gains   \\ \hline
1   & 18.42     & 13.13(\textbf{29\%})     & 1.1     & 185     & 61(\textbf{67\%})      & 6      & 0.11                  & 1127.27 \\
2   & 5.67      & 4.77(\textbf{16\%})      & 0.9     & 380     & 315(\textbf{17\%})     & 5      & 0.08                  & 812.5   \\
3   & 2.12      & 0.94(\textbf{56\%})      & 1.1     & 362     & 290(\textbf{20\%})     & 23     & 0.23                  & 311.69  \\
4   & 3.1       & 2.98(\textbf{4\%})      & 0.9     & 231     & 126(\textbf{45\%})     & 6      & 0.13                  & 807.69  \\
5   & 3.52      & 2.63(\textbf{25\%})      & 3.6     & 627     & 187(\textbf{70\%})     & 39     & 1.11                  & 395.33  \\
6   & 44.78     & 13.23(\textbf{70\%})     & 2.2     & 2087    & 869(\textbf{58\%})     & 35     & 0.67                  & 1828.83 \\
7   & 17.17     & 12.95(\textbf{25\%})     & 8.8     & 1435    & 524(\textbf{63\%})     & 423    & 3.34                  & 272.67  \\
8   & 79.98     & 74.89(\textbf{6\%})     & 7.9     & 3795    & 3485(\textbf{8\%})    & 118    & 2.23                  & 139.14  \\
9   & 102.7     & 77.9(\textbf{24\%})      & 19.7    & 8075    & 5641(\textbf{30\%})    & 312    & 8.67                  & 280.9   \\
10  & 2.52      & 2.08(\textbf{17\%})      & 11      & 409     & 343(\textbf{16\%})     & 65     & 0.14                  & 471.43  \\
11  & 18        & 7.32(\textbf{59\%})      & 32.6    & 10281   & 3450(\textbf{66\%})    & 497    & 14.22                 & 480.48  \\
12  & 32.06     & 18.83(\textbf{41\%})     & 10.6    & 5172    & 1689(\textbf{67\%})    & 114    & 4.91                  & 709.95  \\
13  & 215.11    & 178.11(\textbf{17\%})    & 35.7    & 19484   & 16569(\textbf{15\%})   & 515    & 15.75                 & 185.07  \\
14  & 336.37     & 234.88(\textbf{30\%})     & 21.9    & 10848   & 4454(\textbf{59\%})    & 536    & 12.11                 & 527.91  \\ \hline
15  & 131.92     & 83.77(\textbf{36\%})     & 12.4    & 8261    & 3660(\textbf{56\%})    & 124    & 4.01                  & 1147.95 \\
16  & 144.93     & 97.82(\textbf{33\%})     & 22.7    & 11018   & 4710(\textbf{57\%})    & 348    & 8.1                   & 778.48  \\
17  & 393.99     & 269.69(\textbf{32\%})     & 36.3    & 16157   & 8050(\textbf{50\%})    & 471    & 17.31                 & 468.48  \\ \hline
avg & 91.31     & 66.65(\textbf{27\%})     & 13.5    & 5812.18 & 3201.35(\textbf{45\%}) & 213.94 & 5.48                  & 632.1   \\ \hline
\end{tabular}
}
\end{table}

In the second experiment, we run Light-Fusion using the same configuration as Fusion to produce $S(V_{\text{src}}, V_{\text{sink}})$. 
Light-Fusion, however, incorporates an additional identification algorithm, CI, to filter out non-contributing summaries before the summary collection and cloning.
We count the number of identified non-contributing summaries and then compare it with the total number of non-contributing summaries in the first experiment to determine the identification ratio.
The findings are listed in the ``\#Identified'' column of Table~\ref{table:benchmark}.

Throughout both experiments, 
the PDG of each benchmark is pre-built once and persisted to disk.
Both Fusion and Light-Fusion read the same PDG as input when analyzing a benchmark.
We monitor the execution time, the memory consumption, and the analysis results, $S(V_{\text{src}}, V_{\text{sink}})$. 
The performance data for the two runnings and the building of PDG are listed in Table~\ref{table:compare}. 
In the second set, we specifically record the time consumed by CI, as the memory usage generally stays low, below 200 MB in 15 benchmarks, with the exceptions of \textit{blender} at 265 MB and \textit{mysqld} at 252 MB.
The running time of CI is listed in the ``CI'' column of Table~\ref{table:compare}.

\textbf{Soundness.}
We compare the analysis results $S(V_{\text{src}}, V_{\text{sink}})$ across both sets and the unsafe sinks reported by Fusion and Light-Fusion in Section~\ref{sec:rq3}. 
They remain the same, demonstrating the preservation of the precision of our approach.

\textbf{Effectiveness.}
While using normal graph reachability (\texttt{bfs}) to collect necessary vertices $V^{N}$, CI soundly approximates the summary contribution without considering context sensitivity. 
Thus, some redundant summaries could be incorrectly classified as contributing ones. 
Column ``\#Identified'' in Table~\ref{table:benchmark} lists the number of redundant summaries that can be identified and the percentage relative to redundant summaries that occur for each benchmark.
We observe that CI can precisely identify redundant summaries, reaching the high ratio from 62\% (in \textit{ffmpeg}) to 91\% (in \textit{perlbench}). 
On average, CI correctly identifies 79\% of redundant summaries.
One could reject the bogus $V^{N}$ by reaching context sensitivity.
However, handling context sensitivity may be more costly. 
We examine this aspect in Section~\ref{sec:rq2}. 
The high rate of identification achieved by CI in most projects shows that there is limited room for improvement with a context-sensitive identification algorithm.

\textbf{Efficiency.}
Table~\ref{table:compare} presents the running performance of Fusion, Light-Fusion, CI, and the performance gains when applying CI.
Light-Fusion, using our CI to prune redundant summaries, includes CI's running time in its performance metrics.
For all benchmarks, CI's running time for collecting $V^{N}$ remains below a minute.
Specifically, using CI significantly improves the running time and memory of Light-Fusion, as evidenced by the data in the ``Time'' and ``Memory'' columns of the table.
Without computing and maintaining the redundant summaries captured by CI, 
the average running time of Fusion is reduced from 5812.18 seconds to 3201.35 seconds, saving over 40 minutes (or 2,610.83 seconds).
Also, the average memory usage for Fusion drops from 44.26 GB to 33.88 GB.

\begin{table}[] \small
\caption{
\texttt{\#Solver} is the number of saved solver calls, which is equal to the number of identified redundant summaries (\texttt{\#Identified} column in Table~\ref{table:benchmark}).
The total (\texttt{T}) reduction in performance,
performance of calling the solver (\texttt{S}), with the percentage of each with respect to the total reduction in performance shown in parentheses.
}
\label{table:breakdown}
\centering

\resizebox{0.8\columnwidth}{!} {

\begin{tabular}{cr|rr|rr}
\hline
\multicolumn{1}{l}{\multirow{2}{*}{ID}} & \multicolumn{1}{l|}{\multirow{2}{*}{\#Solver}} & \multicolumn{2}{c|}{Time(s)}                   & \multicolumn{2}{c}{Memory(GB)}                \\
\multicolumn{1}{l}{}                    & \multicolumn{1}{l|}{}                          & \multicolumn{1}{c}{T} & \multicolumn{1}{c|}{S} & \multicolumn{1}{c}{T} & \multicolumn{1}{c}{S} \\ \hline
\multicolumn{1}{c|}{1}                  & 9.3K                                           & 124                   & 110(\textbf{89\%})     & 5.29                  & 1.3(\textbf{25\%})    \\
\multicolumn{1}{c|}{2}                  & 3.7K                                           & 65                    & 59(\textbf{91\%})      & 0.9                   & 0.1(\textbf{11\%})    \\
\multicolumn{1}{c|}{3}                  & 19.0K                                          & 72                    & 66(\textbf{92\%})      & 1.18                  & 0.26(\textbf{22\%})   \\
\multicolumn{1}{c|}{4}                  & 8.4K                                           & 105                   & 95(\textbf{90\%})      & 0.12                  & 0.01(\textbf{8\%})    \\
\multicolumn{1}{c|}{5}                  & 66.2K                                          & 440                   & 403(\textbf{92\%})     & 0.89                  & 0.18(\textbf{20\%})   \\
\multicolumn{1}{c|}{7}                  & 257.5K                                         & 911                   & 840(\textbf{92\%})     & 4.22                  & 1(\textbf{24\%})      \\
\multicolumn{1}{c|}{8}                  & 22.5K                                          & 310                   & 280(\textbf{90\%})     & 5.09                  & 1.4(\textbf{28\%})    \\
\multicolumn{1}{c|}{9}                  & 197.7K                                         & 2434                  & 2222(\textbf{91\%})    & 24.8                  & 5.6(\textbf{23\%})    \\
\multicolumn{1}{c|}{10}                 & 7.2K                                           & 66                    & 59(\textbf{89\%})      & 0.44                  & 0.1(\textbf{23\%})    \\
\multicolumn{1}{c|}{11}                 & 555.9K                                         & 6831                  & 6403(\textbf{94\%})    & 10.68                 & 3.3(\textbf{31\%})    \\
\multicolumn{1}{c|}{12}                 & 258.9K                                         & 3483                  & 3320(\textbf{95\%})    & 13.23                 & 4.3(\textbf{33\%})    \\
\multicolumn{1}{c|}{13}                 & 287.4K                                         & 2915                  & 2817(\textbf{97\%})    & 37                    & 8.9(\textbf{24\%})    \\
\multicolumn{1}{c|}{14}                 & 505.6K                                         & 6394                  & 6188(\textbf{97\%})    & 101.49                & 37.1(\textbf{37\%})   \\ \hline
\multicolumn{1}{c|}{15}                 & 165.8K                                         & 4601                  & 4403(\textbf{96\%})    & 48.15                 & 12(\textbf{25\%})     \\
\multicolumn{1}{c|}{16}                 & 244.7K                                         & 6308                  & 6099(\textbf{97\%})    & 47.11                 & 8.8(\textbf{19\%})    \\
\multicolumn{1}{c|}{17}                 & 519.2K                                         & 8107                  & 7803(\textbf{96\%})    & 124.3                 & 26.8(\textbf{22\%})   \\ \hline
\multicolumn{1}{c|}{avg}                & 268.7K                                         & 2610                  & 2487(\textbf{95\%})    & 24.7                  & 6.66(\textbf{27\%})   \\ \hline
\end{tabular}

}
\vspace{-0.4cm}
\end{table}

\textbf{Breakdown.}
To investigate how much resource is spent collecting redundant summaries and how much resource is spent on calling the constraint solver to verify the conditions of redundant summaries, we break down the reduced performance in Table~\ref{table:compare}. 
This is obtained by computing the difference in running time and memory between Fusion and Light-Fusion, as shown in Table~\ref{table:compare}. 
Additionally, we present the number of solver calls avoided in Table~\ref{table:breakdown}.
Once the summary path is collected, the constraint solver is called; thus, the avoided calls of the constraint solver equal the number of redundant summaries identified, 
as shown in the data in the columns ``\#Identified'' in Table~\ref{table:benchmark} and ``\#Solver'' in Table~\ref{table:breakdown}.
For the total reduced running time, almost 90\% is spent on verifying the conditions of redundant summaries across the benchmarks. 
As the size of the program grows, the complexity of the summary conditions tends to increase as well. 
This results in longer summary paths and more time required to solve these conditions. 
Therefore, the proportion of time spent on the solver increases from 89\% to 97\%.
On the other hand, for the total reduced memory, almost 73\% is consumed by storing the collected summaries.

\textbf{Performance Gains.}
It is noted that the running time of CI does not hurt the performance of Light-Fusion. 
The ``Gains'' column in the table illustrates the performance gains achieved by using CI, which is the ratio of the time saved by Light-Fusion to the overhead incurred by CI. 
In the \textit{povray} (ID 6) case, CI achieves the highest performance gain of 1828.83 $\times$ by reducing 1218 seconds with only 0.67 seconds of overhead.
In the \textit{mysqld} (ID 17) case, CI reduces the most time, saving 8107 seconds (2.25 hours), which is nearly half of the original Fusion running time, with only 17.31 seconds of overhead.
On average, CI achieves a performance gain of 632.1$\times$.



\subsection{RQ2: Performance Boosting}
\label{sec:rq2}


\textbf{Fusion vs. Light-Fusion.}
The performance comparison between Fusion and Light-Fusion is listed in Table~\ref{table:compare}.
The percentage numbers in parentheses represent the extra performance requirements of Fusion against Light-Fusion.
On average, both the time and memory could be reduced by 45\% and by 27\% by using CI.
The time reduction is more significant than the memory reduction since analyzers allocate considerable CPU resources to summary collection and solving path conditions, which can be NP-hard~\cite{fan2019smoke, cook2023complexity}. 
Thus, pruning redundant summaries can save significant time by conserving CPU resources.
The memory reduction percentages are constrained by the number of redundant summaries in the benchmarks, as indicated by the ``\#Redun'' column in Table~\ref{table:benchmark}, which shows an average redundancy ratio of 19\% closely corresponding to the average memory reduction percentages.

\textbf{Light-Fusion vs. CFL-Light-Fusion.}
To explore the impact of enhancing the CI precision, such as by using the CFL reachability~\cite{kodumal2004set, chaudhuri2008subcubic, yannakakis1990graph, melski2000interconvertibility} to reach context sensitivity and identify more redundant summaries compared to \textit{bfs}, we replaced \textit{bfs} in CI with the state-of-the-art CFL reachability algorithm~\cite{lei2022taming}. 
The modified algorithm was executed as the CFL-Light-Fusion instance compared with Fusion and Light-Fusion.
We monitored the running time on all three instances.


\begin{figure}[t]
    \centering
    \includegraphics[width=0.8\columnwidth]{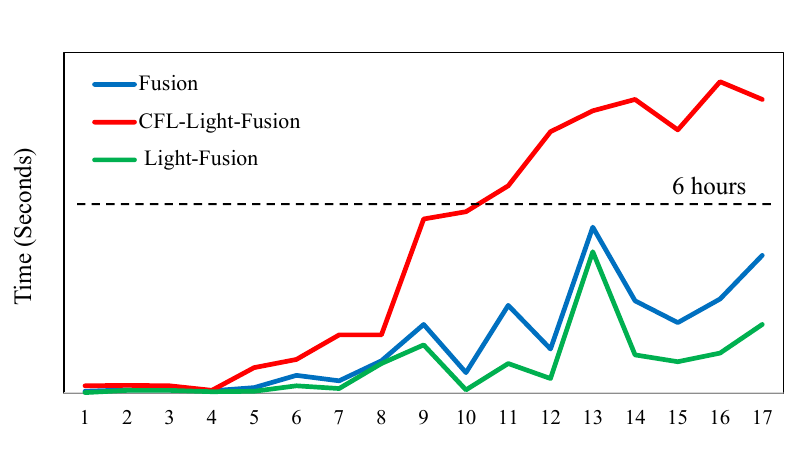}
    \caption{Light-Fusion vs. its variants}
    \label{fig:cfl-light-fusion}
    \vspace{-0.2cm}
\end{figure}

The results presented in Fig.~\ref{fig:cfl-light-fusion} demonstrated that CFL-Light-Fusion did not yield performance improvements; instead, it significantly slowed down the process, particularly as the program size increased. 
Notably, CFL-Light-Fusion failed to analyze benchmarks beyond \textit{cam4} (ID 10) within a six-hour timeframe. 
This decrease in performance can be attributed mainly to the cubic complexity of CFL-based approaches~\cite{kodumal2004set, chaudhuri2008subcubic, yannakakis1990graph, melski2000interconvertibility}, which introduce substantial overhead that outweighs the precision benefits. 
As a result, trading efficiency for precision becomes impractical. 

\subsection{RQ3: Comparing to Top-down Approaches}

\label{sec:rq3}

To evaluate the performance improvements of the path-sensitive analyzer compared to top-down approaches~\cite{reps1995precise, reps1994speeding, arzt2014flowdroid, murphy1999program, sagiv1996precise}, we conducted a comparative analysis between Light-Fusion and \textit{PhASAR}~\cite{10.1007/978-3-030-17465-1_22}. 
Using the latest release~\cite{phasar} of \textit{PhASAR} at the time of writing, we configured it to analyze the same source-sink pairs as Light-Fusion to ensure a fair comparison. 

\begin{figure}[t]
    \centering
    \includegraphics[width=\columnwidth]{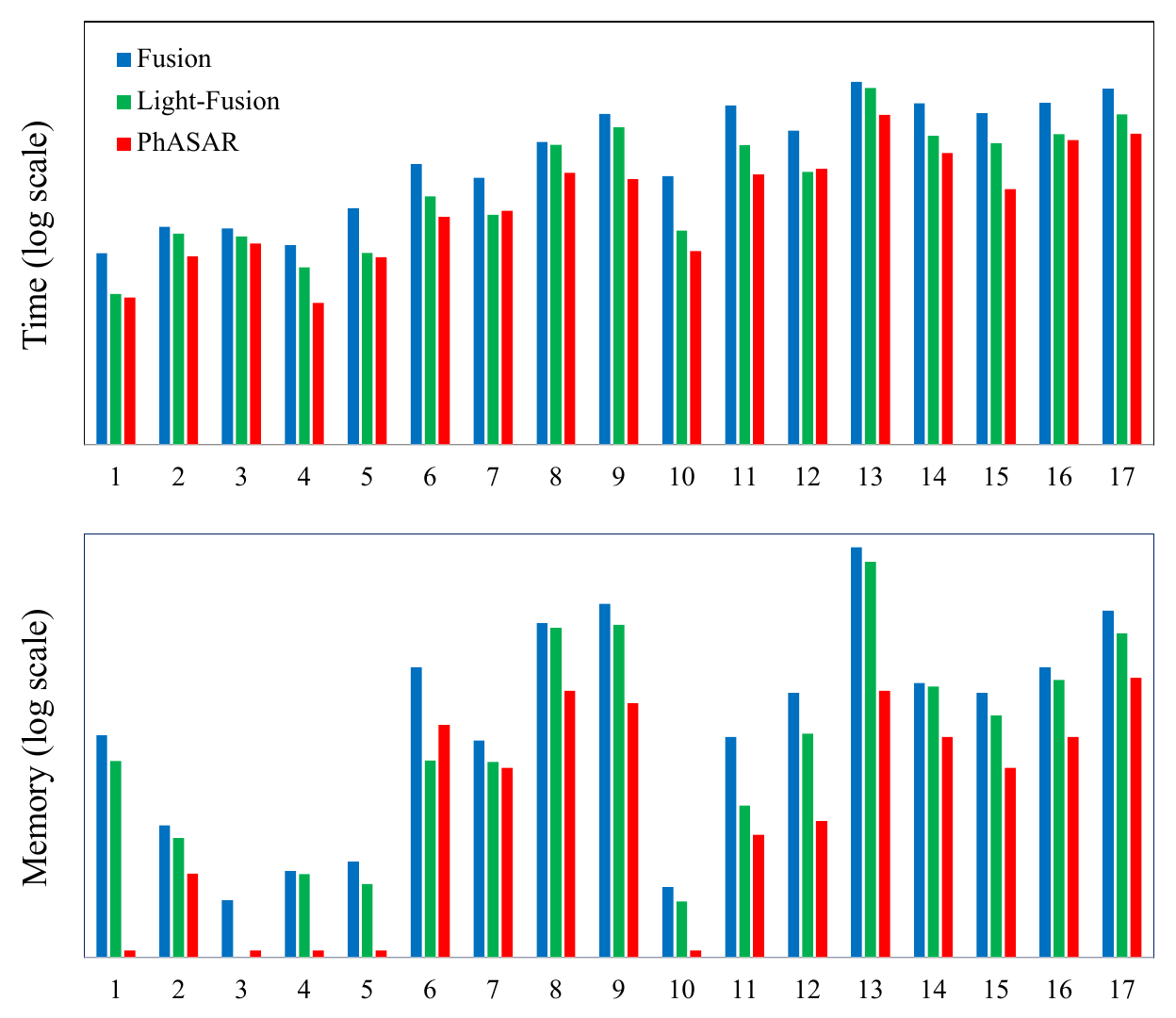}
    \caption{Performance: Fusion vs. Light-Fusion vs. \textit{PhASAR}.}
    \label{fig:td-comparsion}
\end{figure}

\textbf{Performance.}
The results shown in Fig.~\ref{fig:td-comparsion} show the running time and memory usage of Fusion, Light-Fusion, and \textit{PhASAR}, which are represented by blue, green, and red bars, respectively. 
In general, both Fusion and Light-Fusion require more resources compared to \textit{PhASAR}, as indicated by the taller blue and green bars compared to the red bars. 
However, the green bars (representing Light-Fusion) are closer in height to the red bars (\textit{PhASAR}) than the blue bars (representing Fusion). 
In some benchmarks, the green bars are even lower than the red bars. 
For example, the running time of Light-Fusion for \textit{cactus} (ID 7) and \textit{xalanbmk} (ID 12), as well as the memory usage for \textit{x264} (ID 3), are lower than those of \textit{PhASAR}.


Next, we quantify the amount of running time and memory among Fusion, Light-Fusion, and \textit{PhASAR}. 
The results show that Fusion requires 3.35$\times$ the time and 9.75$\times$ the memory consumed by \textit{PhASAR} on average.
However, Light-Fusion can reduce the requirements to just 1.4$\times$ the time and 6.95$\times$ the memory.
Additionally, Light-Fusion has performance comparable with \textit{PhASAR} on benchmarks such as \textit{leea} (ID 1), x264 (ID 3), \textit{omnetpp} (ID 5), \textit{cactus} (ID 7), \textit{xalanbmk} (ID 12), and \textit{ffmpeg} (ID 16), requiring only an additional 0.5$\times$ of \textit{PhASAR}'s execution time. 
However, Fusion could only complete \textit{x264} (ID 3) within the same threshold.

\textbf{Analysis Results.}
We record the unsafe sinks reported by Fusion, Light-Fusion, and \textit{PhASAR} for each benchmark. 
A sink is considered unsafe if there is a feasible path from a source to the sink.
Note that top-down approaches, such as \textit{PhASAR}, usually sacrifice path sensitivity for better scalability. As a result, they may misclassify safe sinks as unsafe ones. 

First, the results show that the unsafe sinks reported by Light-Fusion are the same as those reported by Fusion, demonstrating our approach's precision-preserving nature. 
Next, we examine the unsafe sinks reported by Light-Fusion and \textit{PhASAR}. 
Light-Fusion identifies 90\% of the sinks, which are reported as unsafe by \textit{PhASAR}, as actually being safe, indicating a high false-positive rate for \textit{PhASAR}. 

Additionally, nearly 92\% of unsafe sinks identified by Light-Fusion are also identified by \textit{PhASAR}. 
However, \textit{PhASAR} fails to report some unsafe sinks identified by Light-Fusion due to not modeling the value flow of library functions after manual verification.
We provide the complete comparison data in Appendix Section~\ref{app:comparison}.

In conclusion, our method incorporates path-sensitive analyses without compromising precision and still achieves good performance. 
This aligns with common industrial requirements~\cite{bessey2010few, mcpeak2013scalable} of maintaining both high performance and low false positive rates.

\section{Related Work}
\textbf{Compositional Analysis.}
Many compositional analyses aim to improve efficiency by reusing information within a procedure as summaries~\cite{aikenSaturnManual, babic2008calysto, mcpeak2013scalable, shi2018pinpoint, shi2020pipelining, reps1995precise, sagiv1996precise}, which include top-down and bottom-up summaries.
Most program analyses prefer a bottom-up approach~\cite{aikenSaturnManual, babic2008calysto, mcpeak2013scalable, shi2018pinpoint, shi2020pipelining}. 
In these approaches, a function's effect is represented using bottom-up summaries, and the summary of a callee is inlined into the summary of its caller. 
This avoids redundant analysis of individual functions.
However, when collecting the summary of a callee, it is challenging to determine whether the summary will be useful to callers since the callers are typically analyzed afterward. 
Our work employs a lightweight analysis to pre-compute a set of vertices, which allows us to determine whether a summary should be computed and reduces unnecessary computations.

\textbf{Value Flow Analysis.}
Cherem et al.\cite{cherem2007practical} utilized value flow analysis to detect software bugs such as memory leaks. 
Subsequently, several works have aimed to refine the recall and precision of this analysis\cite{shi2018pinpoint, shi2020conquering, sui2016svf, sui2014detecting, shi2021path}. 
However, most of these analyses are not inter-procedurally path-sensitive, with the exceptions of Pinpoint~\cite{shi2018pinpoint} and Fusion~\cite{shi2021path}.
Fusion is an optimization of the performance issues identified in Pinpoint, which were caused by the explosion of summaries and paths in inter-procedural analysis. Fusion addresses this problem by eliminating the storage of path conditions. 
However, it is worth noting that even with Fusion, there are still redundant summaries and paths being computed in function summaries, which cannot be fully optimized.
While this work helps Fusion overcome redundancy issues related to function summaries.



\section{Conclusion}
We present the contribution identification algorithm, which addresses the redundant summary deficiency in the prior value flow analysis. 
It identifies redundant summaries efficiently and effectively without compromising soundness or efficiency.
Furthermore, it results in an average decrease in the time and memory overhead of state-of-the-art path-sensitive value flow analysis by 45\% and 27\%, respectively.
In the end, it enables path-sensitive analyses without compromising precision and still achieves good performance, making it comparable to path-insensitive analyses.

\section{Acknowledgment}
We thank the anonymous reviewers for valuable feedback on earlier drafts of this paper, which helped improve its presentation.
This work is funded by research donations from Huawei, TCL, and Tencent. Yuandao Cai is the corresponding author.


\balance
\bibliographystyle{IEEEtran}
\bibliography{sigproc}

\appendix

\begin{figure*}[t]
    \centering
    \includegraphics[width=\textwidth]{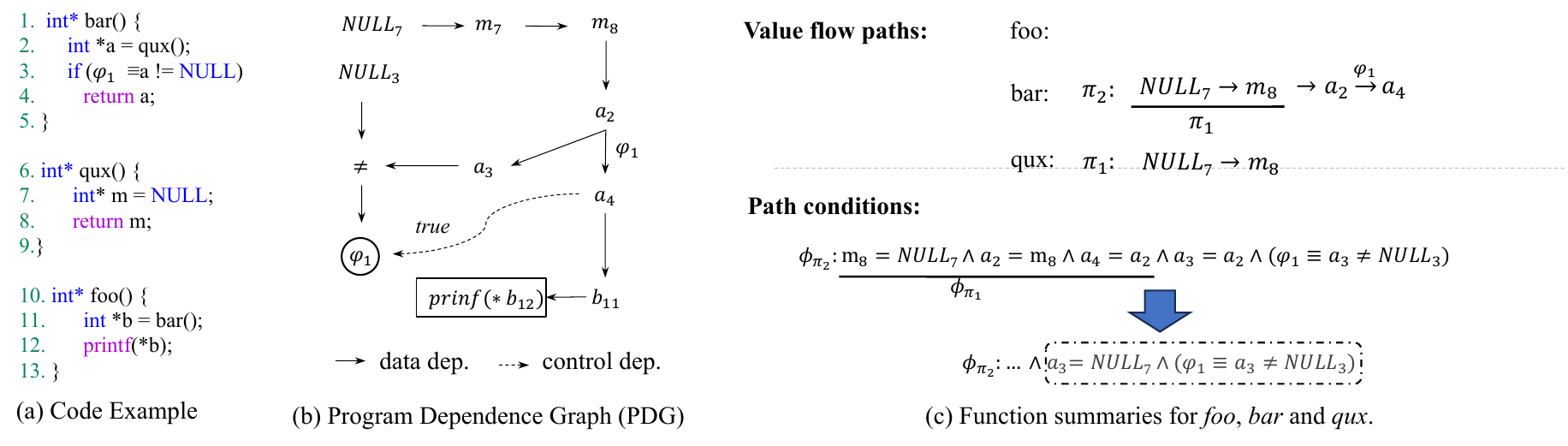}
    \caption{
    An illustration example where the summary $\pi_{2}$ is removed due to the unsatisfiable summary condition $\phi_{\pi_{2}}$.
    Bottom-up analysis for the code shown in (a).
    The (b) shows the corresponding program dependence graph (PDG). 
    (c) shows the function summaries collected during the bottom-up analysis.}
    \label{fig:additional_ex}
    \vspace{-0.4cm}
\end{figure*}

The appendix includes an example that is removed because of the unsatisfiable summary condition in Section~\ref{app:unsat-summary},
soundness proof of contribution identification algorithm~\ref{alg:main}
in Section~\ref{app:sound} and a comparison of results of Fusion, Light-Fusion, and \textit{PhASAR} in Section~\ref{app:comparison}.

\subsection{Pruning Unsatisfied Summary}
\label{app:unsat-summary}
In traditional path-sensitive value flow analysis, if a summary condition is unsatisfiable(\textit{unsat}), it is removed to avoid maintaining an unfeasible summary.
Figure~\ref{fig:additional_ex} provides an example of this.
The summary $\pi_{2}$ that is collected from function ~\textit{bar} will be discarded because its summary condition $\phi_{\pi_{2}}$ is \textit{unsat}.
The summary condition contains clauses indicating that $a_3$ could be a NULL value and not be a NULL value at the same time, as highlighted in the dashed box.
As a result, the function \textit{foo} clones no summary from the function \textit{bar}, and no feasible source-sink path is detected in the case.

\subsection{Soundness Proof of Algorithm~\ref{alg:main}}
\label{app:sound}
\begin{theorem}[Soundness]
\label{thm:sound}

Given $V^{\texttt{N}}$ identified, for any function $f \in P$, if a summary $s=(\pi, \phi)$ is being collected and neither $\pi[0]$ nor $\pi[-1]$ appear in $V^{\texttt{N}}$, it must be a non-contributing summary for function $f$. Canceling the corresponding operations does not affect $S(V_{\text{src}}, V_{\text{sink}})$.
\begin{proof}

Proof by contradiction.

In Algorithm 1, suppose the summary $s$ that is collected between $V_{h}^{f}$ and $V_{t}^{f}$ is a contributing summary, even though $\pi[0] \notin V^{\texttt{N}}$ and $\pi[-1] \notin V^{\texttt{N}}$. 
Thus, the source-sink pair $(src, sink)$ or a guard vertex $g$ labeled on the source-sink path that reaches the head $\pi[0]$ or tail $\pi[-1]$ are not in $V^{\texttt{N}}$. 
However, if such a head or tail vertex exists, Algorithm 2 would include it in $V^{\texttt{N}}$: either on Line 12 in the procedure \texttt{identifyPathContrib}, or on Line 16 in the procedure \texttt{identifyCondContrib}.
Therefore, we arrive at a contradiction, as this implies that $\pi[0] \in V^{\texttt{N}}$ or $\pi[-1] \in V^{\texttt{N}}$. 

In Algorithm 1, when collecting the summary $s$ by concatenating it with the summaries that are cloned from callee functions, following the same judgment conditions as described in the previous case, if it is determined to be a non-contributing summary, the operation of cloning $s^{c}$ should be canceled.
\end{proof}
\end{theorem}

\subsection{Comparison Results of Fusion, Light-Fusion, and \textit{PhASAR}}
\label{app:comparison}

Table~\ref{table:report} shows the classification results of unsafe sinks and sinks reported by Fusion, Light-Fusion, and \textit{PhASAR},
by examining unsafe sinks and safe sinks reported by both tools (LF-US$\cap$P-US and LF-S$\cap$P-S columns), and unsafe sinks reported by one tool but not the other (LF-US$\cap$P-S and LF-S$\cap$P-US columns).

First, the unsafe sinks reported by Fusion (F-US) and Light-Fusion (LF-US) are identical, demonstrating the precision-preserving nature of our approach.

\begin{table}[t] \small
\renewcommand\thetable{A.1}
\caption{Classification of unsafe sinks (US) and safe sinks (S) reported by Fusion (F), Light-Fusion (LF), and \textit{PhASAR} (P).
}
\label{table:report}
\resizebox{\columnwidth}{!} {
\begin{tabular}{r|rccr|crcc}
\hline
ID & \multicolumn{1}{c}{\#Sinks} & \multicolumn{1}{c}{F-US} & LF-US & \multicolumn{1}{c|}{P-US} & \multicolumn{1}{c}{LF-US$\cap$P-US} & \multicolumn{1}{c}{LF-S$\cap$P-S} & \multicolumn{1}{c}{LF-S$\cap$P-US} & \multicolumn{1}{c}{LF-US$\cap$P-S} \\ \hline
1  & 3184                        & 2                        & 2     & 11                        & 2                                    & 3173                               & 9                                   & 0                                   \\
2  & 3581                        & 9                        & 9     & 86                        & 9                                    & 3,495                              & 77                                  & 0                                   \\
3  & 7657                        & 0                        & 0     & 5                         & 0                                    & 7652                               & 5                                   & 0                                   \\
4  & 4273                        & 1                        & 1     & 261                       & 0                                    & 4012                               & 261                                 & 1                                   \\
5  & 27489                       & 28                       & 28    & 207                       & 21                                   & 27282                              & 186                                 & 7                                   \\
6  & 14490                       & 31                       & 31    & 93                        & 24                                   & 14397                              & 69                                  & 7                                   \\
7  & 49769                       & 26                       & 26    & 378                       & 24                                   & 49391                              & 354                                 & 2                                   \\
8  & 12156                       & 12                       & 12    & 111                       & 10                                   & 12045                              & 101                                 & 2                                   \\
9  & 40888                       & 19                       & 19    & 669                       & 16                                   & 40219                              & 653                                 & 3                                   \\
10 & 3418                        & 8                        & 8     & 217                       & 4                                    & 3201                               & 213                                 & 4                                   \\
11 & 215747                      & 35                       & 35    & 103                       & 35                                   & 215644                             & 68                                  & 0                                   \\
12 & 77705                       & 28                       & 28    & 281                       & 24                                   & 77424                              & 257                                 & 4                                   \\
13 & 146208                      & 79                       & 79    & 3592                      & 76                                   & 142616                             & 3516                                & 3                                   \\
14 & 182621                      & 65                       & 65    & 3902                      & 63                                   & 178719                             & 3839                                & 2                                   \\ \hline
15 & 76727                       & 103                      & 103   & 2382                      & 93                                  & 74345                              & 2289                                & 10                                  \\
16 & 146284                      & 92                       & 92    & 1102                      & 84                                   & 145182                             & 1018                                & 8                                   \\
17 & 215431                      & 168                      & 168   & 2007                      & 162                                  & 213424                             & 1845                                & 6                                   \\ \hline
\end{tabular}
}
\vspace{-0.4cm}
\end{table}

Second, as shown in the LF-S$\cap$P-US column, \textit{PhASAR} mistakenly reports a large number of sinks as unsafe, resulting in high false positive rates due to not collecting and verifying the path conditions.
The majority of these sinks actually cannot be successfully reached by sources due to unsatisfiable path conditions.
Different from the Phasar, Light-Fusion considers path conditions, thereby excluding the mistakenly reported sinks from the top-down approaches. 

Additionally, \textit{PhASAR} fails to report some unsafe sinks that are reported by Light-Fusion as shown in the LF-US$\cap$P-S column.

\end{document}